\documentclass[%
 aip,
 amsmath,amssymb,
,reprint,
floatfix
]{revtex4-1}

\usepackage{graphicx}
\usepackage{bm}
\usepackage[utf8]{inputenc}
\usepackage[T1]{fontenc}
\usepackage{mathptmx}
\usepackage{float}
\usepackage{xr}
\makeatletter

\begin{document}

\preprint{AIP/123-QED}

\title{Direct observation of topological charge impacting skyrmion bubble stability in Pt/Ni/Co asymmetric superlattices}

\author{Nisrit Pandey}
    \email{npandey@andrew.cmu.edu}
    
\author{Maxwell Li}
    \affiliation{Department of Materials Science and Engineering, Carnegie Mellon University, Pittsburgh, PA 15213 USA}
    
\author{Marc De Graef} 
\author{Vincent Sokalski}
    \affiliation{Department of Materials Science and Engineering, Carnegie Mellon University, Pittsburgh, PA 15213 USA}

\date{\today}

\begin{abstract}

We characterize the magnetic properties and domain structure of Pt/Ni/Co asymmetric superlattices in comparison to the more established Pt/Co/Ni structure.  This reversal in stacking sequence leads to a marked drop in interfacial magnetic anisotropy and the magnitude of the interfacial Dzyaloshinskii-Moriya interaction (DMI) as inferred from the DW structure, which we speculate could be related to a degradation of the Pt/Co interface when Pt is deposited on top of the Co layer.  Lorentz transmission electron microscopy reveals exclusively N\'eel type domain walls and, with a perpendicular field, N\'eel skyrmions in the Pt/Co/Ni films.  Conversely, the Pt/Ni/Co samples show only achiral Bloch domain walls, which leads to the formation of achiral Bloch ($Q=1$) and type II bubbles ($Q=0$) at increased perpendicular field.  Combined with the reduced anisotropy leading to greater bubble densities, the latter case makes for an excellent test bed to examine the benefits of topological charge on stability.  Simultaneous observation of Bloch and type II bubbles shows a roughly 50 mT larger annihilation field for the former.  An in-plane component to the magnetic field is shown to both impact the structure of the formed bubbles and separately suppress the topological benefit.
 
\end{abstract}

\maketitle



\begin{figure}[t]
\includegraphics[width = 3.5 in]{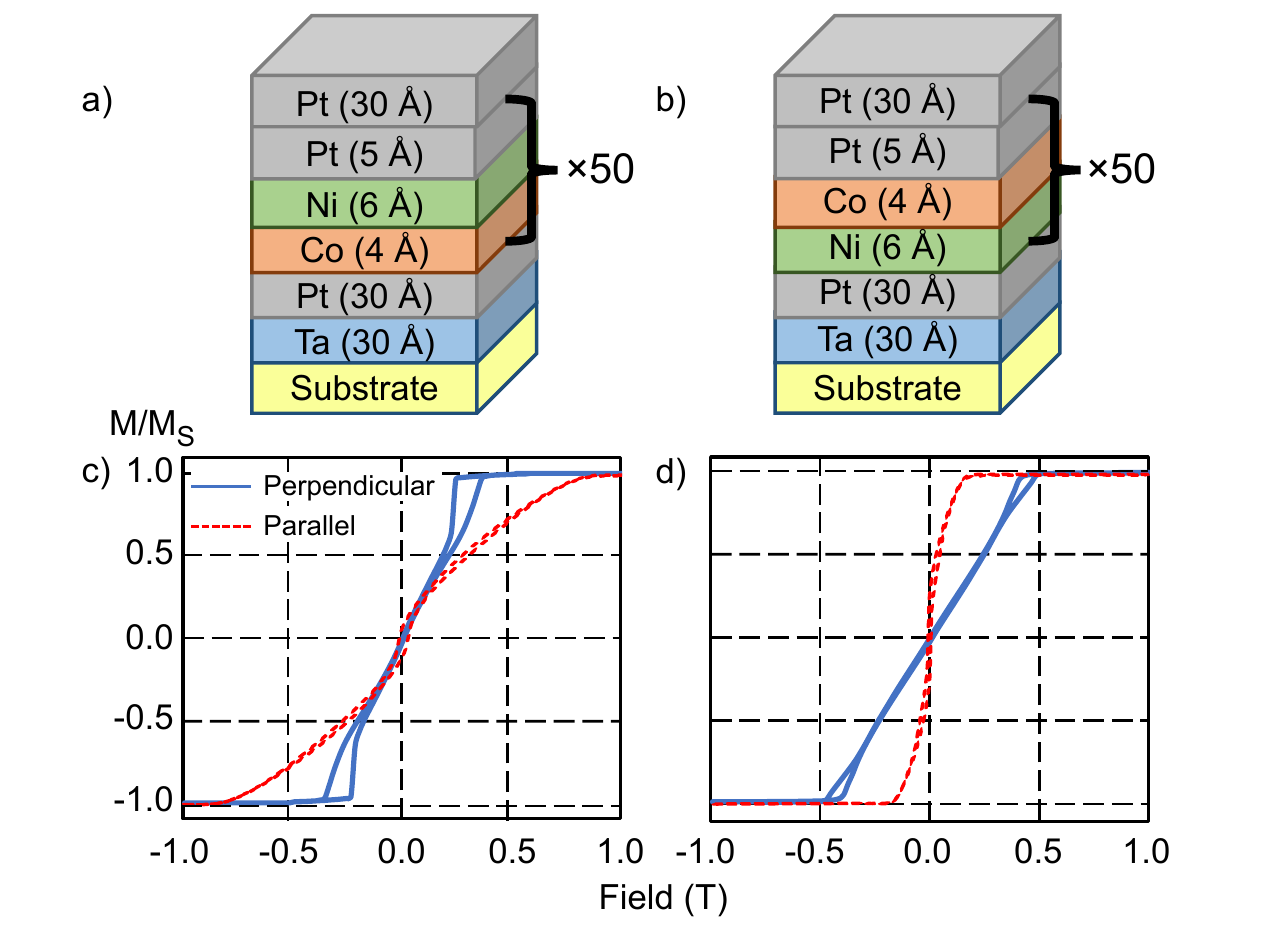}
\caption{\label{Figure1} Schematic stacking sequences of the a) [Pt/Co/Ni]$_{50}$ and b) a) [Pt/Ni/Co]$_{50}$ multi-layers examined in this work. Perpendicular (blue) and parallel (red) M-H loops of c) Pt/Co/Ni and d) Pt/Ni/Co samples showing characteristic bubble shapes and a marked drop in anisotropy for Pt/Ni/Co.}
\end{figure}


\begin{figure*}[t]
\includegraphics[width = 7 in]{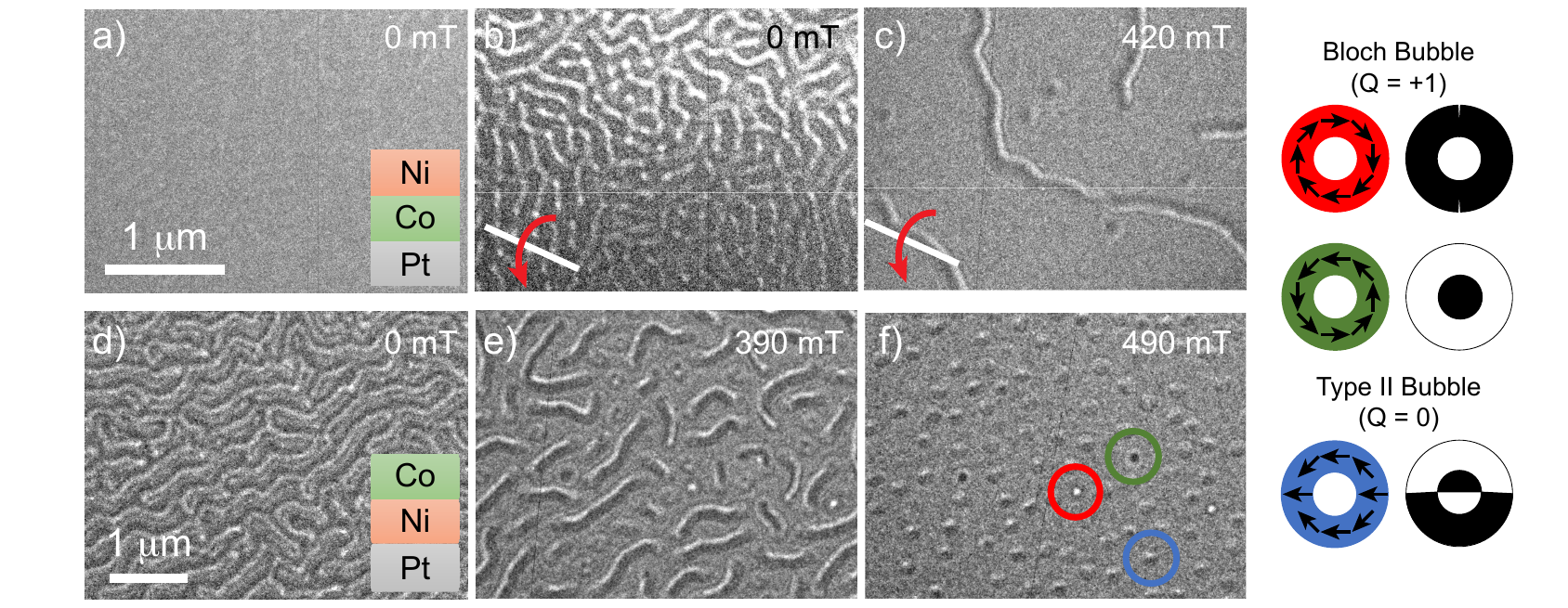}
\caption{\label{Figure2}Fresnel-mode LTEM micrographs of a-c) [Pt/Co/Ni]$_{50}$ and d-f) [Pt/Ni/Co]$_{50}$ under various tilt (indicated by curved arrow) and external field conditions (value inset in figure). a) Zero field, zero tilt condition showing absence of contrast. b) Zero field with $20^\circ$ tilt revealing presence of N\'eel domain walls. c) 420 mT perpendicular magnetic field showing the formation of worm-like domains and N\'eel skyrmions. d) labyrinth Bloch domain structure at zero field/tilt e) transition to an intermediate bubble/worm domain pattern at 390 mT perpendicular field f) achiral Bloch/type 2 bubble array at increased perpendicular field. Example bubble forms are color circled with the corresponding structure, topological charge, and schematic resultant contrast shown in the legend to the right.}
\end{figure*}


The Dzyaloshinskii-Moriya interaction (DMI) found in bulk magnetic non-centrosymmetric crystals\cite{Dzyaloshinsky1958,Moriya1960} and at the heavy metal/ferromagnetic interfaces of magnetic multi-layer films\cite{Thiaville2012} has generated considerable interest in magnetic bubbles with inherent topological protection, also known as skyrmions\cite{muhlbauer2009,Yu2010,Huang2012,Woo2016,Li2019}. These spin structures have been proposed for a wide range of spintronic applications, including devices for next-generation data storage\cite{Kiselev2011,Tomasello2014} and neuromorphic computing \cite{Huang2017,Song2020} as they can be manipulated with remarkable efficiency using magnetic fields and electric currents\cite{Fert2013,Juge2019}.  The bubble/skyrmion topology is defined by a topological charge $Q=\frac{1}{4\pi}\int\mathbf{m}\cdot(\partial_x\mathbf{m} \times \partial_y\mathbf{m})\mathrm{d}x\mathrm{d}y$, where $\mathbf{m}$ is the unit magnetization vector\cite{Nagaosa2013}. A strong interfacial DMI supports the formation of N\'eel skyrmions or domain walls (DWs) where their chirality is influenced by the sign of DMI.  In the case of weak DMI, achiral Bloch DWs and bubbles are observed.  Additionally, topologically trivial bubbles ($Q=0$), also known as type II bubbles, have been observed in the case of weak DMI alongside achiral Bloch bubbles (see Figure \ref{Figure2} schematic).\cite{Li2019} These type II bubbles contain two vertical Bloch lines (with parallel alignment), which are 180$^\circ$ transitions of the internal magnetization in a Bloch domain wall. It is noteworthy that, because a uniformly magnetized film has $Q=0$, there is no required break in topology to annihilate a type 2 bubbles (i.e. the spins of a type II bubble can be continuously rotated into alignment with the outer domain matrix).  This is generally believed to be the origin of higher field stability in topologically non-trivial bubbles.  However, direct in-situ observation of this difference has not been previously reported and the conditions to selectively stabilize certain bubble types as well as the role on topological protection on annihilation of these structures under magnetic fields remain at issue.

The Pt/Co/Ni multi-layer system has shown promising use in the development of future chiral spintronic devices due to the tunability of its interfacial magnetic properties, such as DMI and perpendicular magnetic anisotropy (PMA) \cite{You2012,Rai2020}. However, the vast majority of the work has focused on the case of a Pt$\rightarrow$Co$\rightarrow$Ni deposition order. Here, the Pt/Co interface is credited with the predominantly left-handed chirality observed and is believed to display a stronger interfacial DMI than that at the Ni/Pt interface\cite{Yang2015,Di2015}. Based on this current understanding of Pt/Co/Ni multi-layers, symmetry considerations suggest that Pt/Ni/Co multi-layers should give rise to right-handed N\'eel domain walls, which would broaden the versatility of this system and has been considered for use in chiral synthetic antiferromagnetic (SAF) heterostructures \cite{Prudnikov2018,Pandey2020}. However, as noted in the forthcoming sections, the result is actually a change in wall structure to achiral Bloch (rather than a reversal of N\'eel chirality) suggesting weaker DMI, which is coupled with decreased perpendicular magnetic anisotopy.  Nonetheless, the system presents a convenient test bed for characterizing behavior of the aforementioned bubble/skyrmion forms.  Here, we characterize this modified material system using Lorentz transmission electron microscopy (LTEM) including an examination of the added topological stability of Bloch bubbles and the role of an external magnetic field.


[Pt(5)/(Co(4)/Ni(6))]$_{50}$ and [Pt(5)/(Ni(6)/Co(4)]$_{50}$ were deposited onto oxidized Si (100) substrates and 10 nm amorphous Si$_3$N$_4$ TEM membranes via dc magnetron sputtering in an Argon plasma with pressure fixed at 2.5 mTorr and a base pressure of $<3.0 \times 10^{-7}$ Torr. All samples had a Ta(30)/Pt(30) adhesion/seed layer to induce FCC (111) texture and were capped with Pt(30). Thicknesses stated in the parentheses are expressed in \AA. Alternating gradient field magnetometry (AGFM) and vibrating sample magnetometry (VSM) were used to obtain magnetic hysteresis (M-H) loops for determination of saturation magnetization ($M_\text{s}$) and in-plane saturation field ($H_\text{k}$). 

Samples were imaged using Fresnel-mode LTEM performed at room temperature using an FEI Tecnai F20 with an accelerating voltage of $200$ kV in Lorentz mode (with the objective lens turned off). This method utilizes the deflection of electrons through the Lorentz force to image magnetic contrast out of the focus plane \cite{degraef2000d}. A perpendicular magnetic field was applied \textit{in situ} by weakly exciting the objective lens of the microscope. All images were taken between $1.6-2.0$ mm overfocus.



\begin{table}[h]

\caption{Summary of observed magnetic characteristics in Pt/Co/Ni and Pt/Ni/Co.}
\label{Table1}
\scalebox{1.2}{%
\begin{tabular}{|c||c|c|}
\hline
Property          & \textbf{[Pt/Co/Ni]$_\text{\textbf{50}}$}  & \textbf{[Pt/Ni/Co]$_\text{\textbf{50}}$}   \\ \hline
$M_s$ (kA/m)     & 923                & 704                 \\ \hline
$H_k$ (mT)       & 850                & 194                 \\ \hline
Wall Chirality   & Left Handed N\'eel & Achiral Bloch \\ \hline
Bubble Density   & Low                & High                \\ \hline
\end{tabular}%
}

\end{table}


\begin{figure*}[t]
\includegraphics[width = 7 in, ]{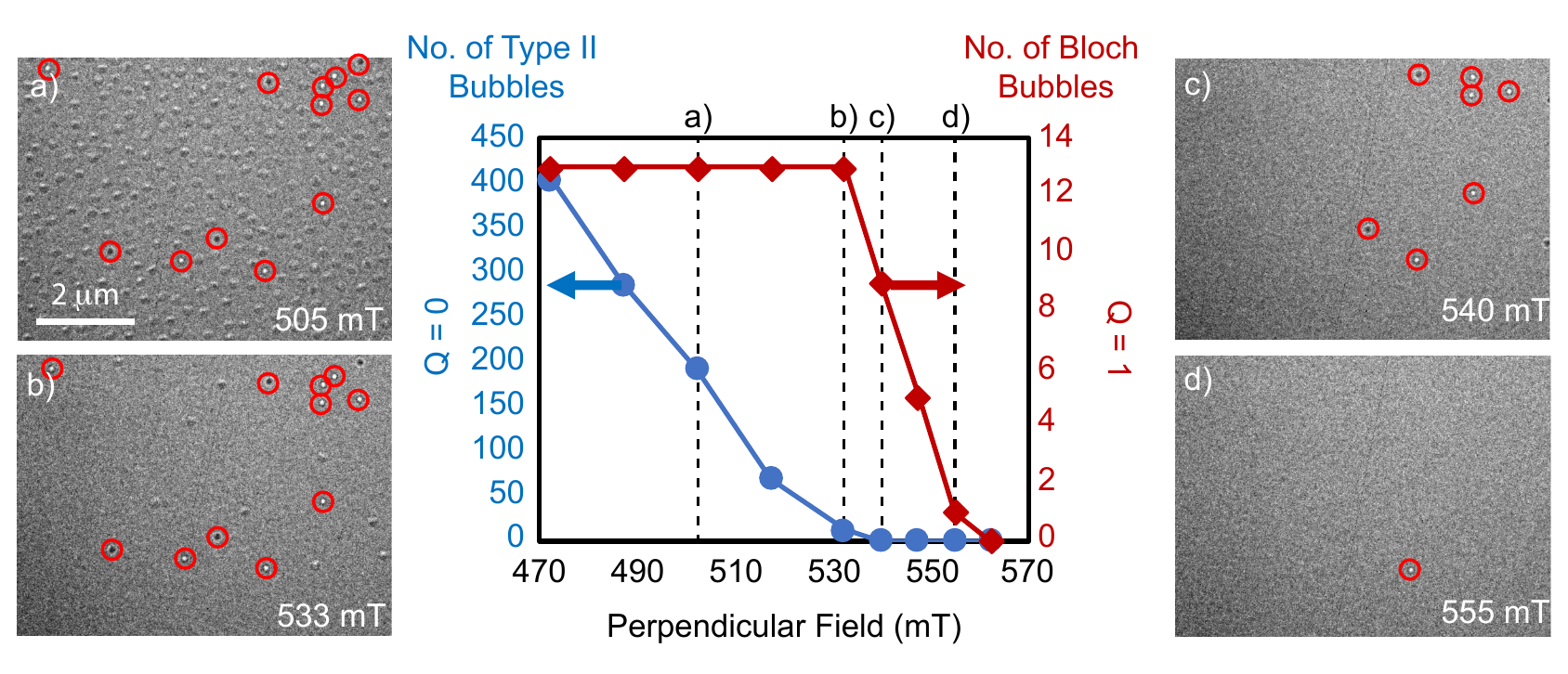}
\caption{\label{Figure3} Center) Tabulated number of Bloch and type 2 bubbles that remain as a function of in-situ perpendicular field in the Pt/Ni/Co sample.  a-d) Example Fresnel-mode LTEM micrographs corresponding to indicated fields on the plot. Bloch bubbles of either chirality ($Q=1$) are highlighted with red circles in each image.  All others are type 2 ($Q=0$).}
\end{figure*}


\begin{figure}[t]
\includegraphics[width = 3 in]{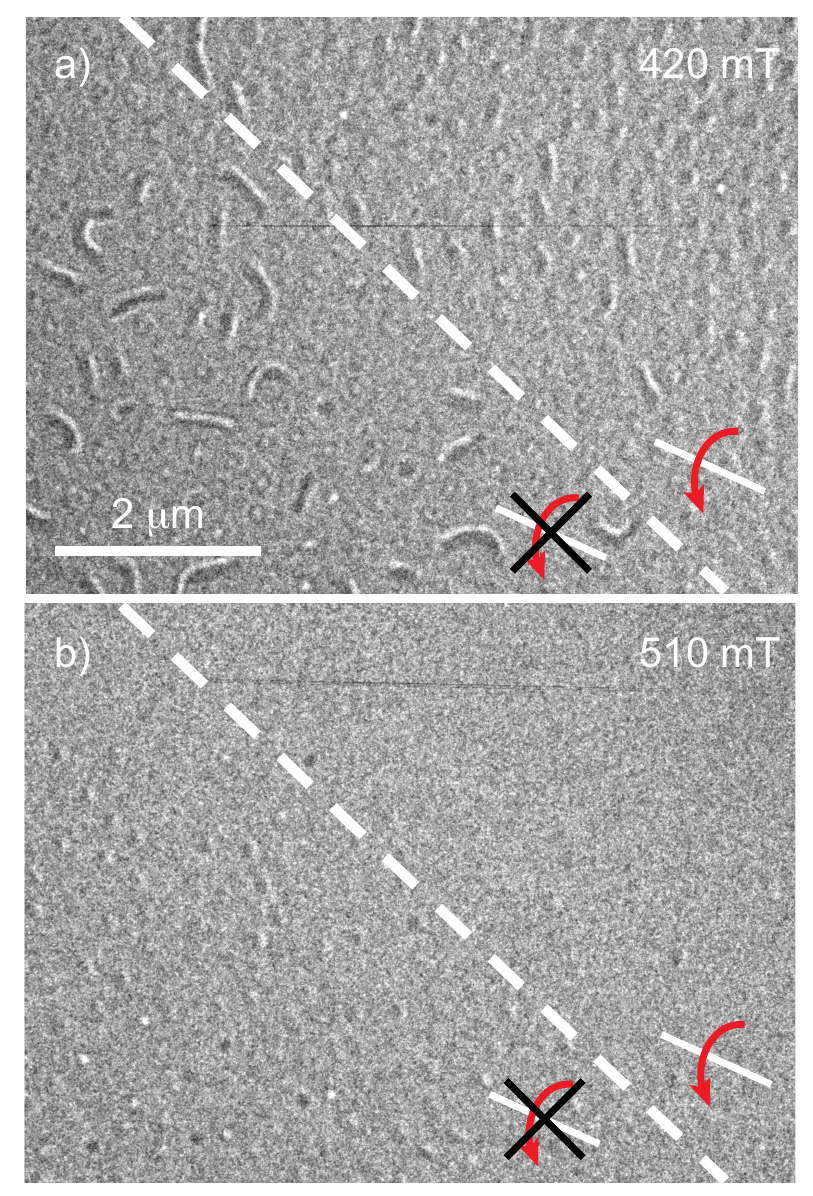}
\caption{\label{Figure4} Fresnel-mode LTEM micrographs of Pt/Ni/Co samples at the location of a bend contour in the TEM membrane. The stability of type II bubble was examined in regions with an inherent tilt (right of dotted line) and without one (left of dotted line). The lack of any type II bubbles remaining in the tilted region (as compared to the many observed in the untilted region) when field is increased form 420 mT to 510 mT suggests the induced, effective in-plane field plays a significant role in their annihilation.} 
\end{figure}

Both [Pt/Co/Ni]$_{50}$ and [Pt/Ni/Co]$_{50}$ multi-layers display typical perpendicular bubble material M-H loops, characterized by a significantly sheared/pinched loop and a sharp nucleation field (Fig.~\ref{Figure1} c,d). Comparing the perpendicular loops of both samples, with all deposition conditions otherwise identical, we find the nucleation field is less pronounced for the case of Pt/Ni/Co. Moreover, the in-plane saturation field and saturation magnetization are comparatively reduced in Pt/Ni/Co (Table \ref{Table1}), which could suggest increased interface diffusion.  Other work has indicated that the deposition order of both Pt/Co and Pd/Co can lead to effective asymmetry in nominally symmetric film stacks\cite{Pollard2017,Lavrijsen2015}.  It is possible that the higher energy Pt adatoms on Co lead to increased diffusion and/or create inhomogeneities compromising the interfacial properties.  

Fresnel-mode LTEM images of [Pt/Co/Ni]$_{50}$ show that domain wall contrast is observed only after an application of a sample tilt (Fig.~\ref{Figure2} a, b), confirming the presence of N\'eel-type DWs and thus an appreciable interfacial DMI to stabilize them. Through the application of a perpendicular field applied \textit{in situ}, we observe the labyrinthine domain pattern (seen at remanence) transition into worm-like domains and eventually $\approx 200$ nm N\'eel skyrmions (Fig.~\ref{Figure2} c).  It is notable that the density of N\'eel skyrmions is relatively small and tend to coexist with worm domains.

Moving towards the [Pt/Ni/Co]$_{50}$ sample, we observe contrast consistent with Bloch domain walls with no apparent preferred chirality (Fig. \ref{Figure2} d). The lack of chiral preference is readily seen at VBLs found along various domain walls where the magnetization flips $180^\circ$. Furthermore, Bloch bubbles of either chirality are observed in the presence of a perpendicular field applied \textit{in situ} differentiated by the core intensity of the observed magnetic contrast. In addition to these achiral bubbles, a large number of type II bubbles are also observed. Further increasing the perpendicular field strength towards saturation, we find that these type II bubbles annihilate before the VBL-free Bloch bubbles do. In fact, despite their initially greater density, we find that every single type II bubble has annihilated while more than half of the Bloch bubbles still remain at ~540 mT. (Fig.~\ref{Figure3}). We attribute the resistance to annihilation of these $Q=1$ bubbles to the energy barrier that must be overcome to transform to a uniformly magnetized state ($Q=1\rightarrow0$).

Having examined the stability of these bubbles in the presence of a perpendicular field, we now consider the effect of in-plane fields, which may be employed for selective nucleation of type II bubbles over skyrmions\cite{Montoya2017,Je2020}, by utilizing intrinsic bends in the underlying TEM membrane. The tilt of these regions produces an effective in-plane field in the presence of a perpendicular field allowing for a direct comparison of its effects as seen in Fig.~\ref{Figure4} \cite{Zhang2018field}. These bends have been reported to be as large as $15^\circ$ with respect to ``flat'' regions of the deposited film/membrane\cite{Fallon2020,Li2020}. With a perpendicular field applied, we observe a larger number of Bloch and type II bubbles in the tilted region as compared to those observed in the ``flat'' region in addition to worm-like domains. When the field is increased further, most, if not all, magnetic contrast is absent in the tilted region suggesting the effective in-plane fields assisted in the annihilation of these bubbles despite being observed to be stable at larger perpendicular fields. Although we postulate that type II bubbles are similarly less stable in the wake of in-plane fields as they are in perpendicular fields when compared to $Q=1$ bubbles, it is interesting that both are ``equally'' afflicted by in-plane fields.


In summary, we have characterized the magnetic properties of [Pt/Ni/Co] multi-layers as compared to the more common [Pt/Co/Ni] structure.  We note a substantial decrease in magnetic anisotropy and saturation magnetization along with a purported decrease in DMI (as inferred from LTEM), which we attribute to increased interdiffusion at the Pt/Co interface.  The [Pt/Ni/Co] multi-layer adopts a domain structure comprised of achiral Bloch DWs as compared to chiral N\'eel-type DWs in the [Pt/Co/Ni] film stack.  With application of a perpendicular field, we observe a relatively dense combination of type II and achiral Bloch bubbles.  A marked difference in the perpendicular annihilation field is found between these bubble types.  Moreover, an in-plane component of the field enables the preferential nucleation of type II bubbles as well as a leveling of the required perpendicular annihilation field for the two bubble types.  Although the properties of the Pt/Ni/Co system are expected to be dependent on processing parameters, the observations here add to the robust tunability of Pt-Co-Ni multi-layer systems and further insight into the role of Pt/Co deposition sequence on magnetic properties.  This particular case also offers a valuable test bed for examining topological stability due to the high density and variety of bubble types present.

\begin{acknowledgments}
The authors acknowledge valuable discussion with Dr. Hans Nembach and Dr. Justin Shaw from the National Institute of Standards and Technology.  This work is financially supported by the Defense Advanced Research Project Agency (DARPA) program on Topological Excitations in Electronics (TEE) under grant number D18AP00011. The authors also acknowledge use of the Materials Characterization Facility at Carnegie Mellon University supported by grant MCF-677785.

\end{acknowledgments}

\section*{Data Availability Statement}
The data that supports the findings of this study are available within the article.

\bibliography{aipsamp}

\begin{thebibliography}{29}%
\makeatletter
\providecommand \@ifxundefined [1]{%
 \@ifx{#1\undefined}
}%
\providecommand \@ifnum [1]{%
 \ifnum #1\expandafter \@firstoftwo
 \else \expandafter \@secondoftwo
 \fi
}%
\providecommand \@ifx [1]{%
 \ifx #1\expandafter \@firstoftwo
 \else \expandafter \@secondoftwo
 \fi
}%
\providecommand \natexlab [1]{#1}%
\providecommand \enquote  [1]{``#1''}%
\providecommand \bibnamefont  [1]{#1}%
\providecommand \bibfnamefont [1]{#1}%
\providecommand \citenamefont [1]{#1}%
\providecommand \href@noop [0]{\@secondoftwo}%
\providecommand \href [0]{\begingroup \@sanitize@url \@href}%
\providecommand \@href[1]{\@@startlink{#1}\@@href}%
\providecommand \@@href[1]{\endgroup#1\@@endlink}%
\providecommand \@sanitize@url [0]{\catcode `\\12\catcode `\$12\catcode
  `\&12\catcode `\#12\catcode `\^12\catcode `\_12\catcode `\%12\relax}%
\providecommand \@@startlink[1]{}%
\providecommand \@@endlink[0]{}%
\providecommand \url  [0]{\begingroup\@sanitize@url \@url }%
\providecommand \@url [1]{\endgroup\@href {#1}{\urlprefix }}%
\providecommand \urlprefix  [0]{URL }%
\providecommand \Eprint [0]{\href }%
\providecommand \doibase [0]{http://dx.doi.org/}%
\providecommand \selectlanguage [0]{\@gobble}%
\providecommand \bibinfo  [0]{\@secondoftwo}%
\providecommand \bibfield  [0]{\@secondoftwo}%
\providecommand \translation [1]{[#1]}%
\providecommand \BibitemOpen [0]{}%
\providecommand \bibitemStop [0]{}%
\providecommand \bibitemNoStop [0]{.\EOS\space}%
\providecommand \EOS [0]{\spacefactor3000\relax}%
\providecommand \BibitemShut  [1]{\csname bibitem#1\endcsname}%
\let\auto@bib@innerbib\@empty
\bibitem [{\citenamefont {Dzyaloshinsky}(1958)}]{Dzyaloshinsky1958}%
  \BibitemOpen
  \bibfield  {author} {\bibinfo {author} {\bibfnamefont {I.}~\bibnamefont
  {Dzyaloshinsky}},\ }\bibfield  {title} {\enquote {\bibinfo {title} {A
  thermodynamic theory of ``weak" ferromagnetism of antiferromagnetics},}\
  }\href {\doibase https://doi.org/10.1016/0022-3697(58)90076-3} {\bibfield
  {journal} {\bibinfo  {journal} {J. Phys. Chem. Solids}\ }\textbf {\bibinfo
  {volume} {4}},\ \bibinfo {pages} {241--255} (\bibinfo {year}
  {1958})}\BibitemShut {NoStop}%
\bibitem [{\citenamefont {Moriya}(1960)}]{Moriya1960}%
  \BibitemOpen
  \bibfield  {author} {\bibinfo {author} {\bibfnamefont {T.}~\bibnamefont
  {Moriya}},\ }\bibfield  {title} {\enquote {\bibinfo {title} {Anisotropic
  superexchange interaction and weak ferromagnetism},}\ }\href {\doibase
  https://doi.org/10.1103/PhysRev.120.91} {\bibfield  {journal} {\bibinfo
  {journal} {Phys. Rev.}\ }\textbf {\bibinfo {volume} {120}},\ \bibinfo {pages}
  {91--98} (\bibinfo {year} {1960})}\BibitemShut {NoStop}%
\bibitem [{\citenamefont {Thiaville}\ \emph {et~al.}(2012)\citenamefont
  {Thiaville}, \citenamefont {Rohart}, \citenamefont {Jue}, \citenamefont
  {Cros},\ and\ \citenamefont {Fert}}]{Thiaville2012}%
  \BibitemOpen
  \bibfield  {author} {\bibinfo {author} {\bibfnamefont {A.}~\bibnamefont
  {Thiaville}}, \bibinfo {author} {\bibfnamefont {S.}~\bibnamefont {Rohart}},
  \bibinfo {author} {\bibfnamefont {E.}~\bibnamefont {Jue}}, \bibinfo {author}
  {\bibfnamefont {V.}~\bibnamefont {Cros}}, \ and\ \bibinfo {author}
  {\bibfnamefont {A.}~\bibnamefont {Fert}},\ }\bibfield  {title} {\enquote
  {\bibinfo {title} {Dynamics of {D}zyaloshinskii domain walls in ultrathin
  magnetic films},}\ }\href {\doibase
  http://dx.doi.org/10.1209/0295-5075/100/57002} {\bibfield  {journal}
  {\bibinfo  {journal} {EPL}\ }\textbf {\bibinfo {volume} {100}} (\bibinfo
  {year} {2012}),\ http://dx.doi.org/10.1209/0295-5075/100/57002}\BibitemShut
  {NoStop}%
\bibitem [{\citenamefont {Muhlbauer}\ \emph {et~al.}(2009)\citenamefont
  {Muhlbauer}, \citenamefont {Binz}, \citenamefont {Jonietz}, \citenamefont
  {Pfleiderer}, \citenamefont {Rosch}, \citenamefont {Neubauer}, \citenamefont
  {Georgii},\ and\ \citenamefont {Boni}}]{muhlbauer2009}%
  \BibitemOpen
  \bibfield  {author} {\bibinfo {author} {\bibfnamefont {S.}~\bibnamefont
  {Muhlbauer}}, \bibinfo {author} {\bibfnamefont {B.}~\bibnamefont {Binz}},
  \bibinfo {author} {\bibfnamefont {F.}~\bibnamefont {Jonietz}}, \bibinfo
  {author} {\bibfnamefont {C.}~\bibnamefont {Pfleiderer}}, \bibinfo {author}
  {\bibfnamefont {A.}~\bibnamefont {Rosch}}, \bibinfo {author} {\bibfnamefont
  {A.}~\bibnamefont {Neubauer}}, \bibinfo {author} {\bibfnamefont
  {R.}~\bibnamefont {Georgii}}, \ and\ \bibinfo {author} {\bibfnamefont
  {P.}~\bibnamefont {Boni}},\ }\bibfield  {title} {\enquote {\bibinfo {title}
  {Skyrmion lattice in a chiral magnet},}\ }\href {\doibase
  10.1126/science.1166767} {\bibfield  {journal} {\bibinfo  {journal}
  {Science}\ }\textbf {\bibinfo {volume} {323}},\ \bibinfo {pages} {915–919}
  (\bibinfo {year} {2009})}\BibitemShut {NoStop}%
\bibitem [{\citenamefont {Yu}\ \emph {et~al.}(2010)\citenamefont {Yu},
  \citenamefont {Onose}, \citenamefont {Kanazawa}, \citenamefont {Park},
  \citenamefont {Han}, \citenamefont {Matsui}, \citenamefont {Nagaosa},\ and\
  \citenamefont {Tokura}}]{Yu2010}%
  \BibitemOpen
  \bibfield  {author} {\bibinfo {author} {\bibfnamefont {X.~Z.}\ \bibnamefont
  {Yu}}, \bibinfo {author} {\bibfnamefont {Y.}~\bibnamefont {Onose}}, \bibinfo
  {author} {\bibfnamefont {N.}~\bibnamefont {Kanazawa}}, \bibinfo {author}
  {\bibfnamefont {J.~H.}\ \bibnamefont {Park}}, \bibinfo {author}
  {\bibfnamefont {J.~H.}\ \bibnamefont {Han}}, \bibinfo {author} {\bibfnamefont
  {Y.}~\bibnamefont {Matsui}}, \bibinfo {author} {\bibfnamefont
  {N.}~\bibnamefont {Nagaosa}}, \ and\ \bibinfo {author} {\bibfnamefont
  {Y.}~\bibnamefont {Tokura}},\ }\bibfield  {title} {\enquote {\bibinfo {title}
  {Real-space observation of a two-dimensional skyrmion crystal},}\ }\href@noop
  {} {\bibfield  {journal} {\bibinfo  {journal} {Nature}\ }\textbf {\bibinfo
  {volume} {465}},\ \bibinfo {pages} {901} (\bibinfo {year}
  {2010})}\BibitemShut {NoStop}%
\bibitem [{\citenamefont {Huang}\ and\ \citenamefont
  {Chien}(2012)}]{Huang2012}%
  \BibitemOpen
  \bibfield  {author} {\bibinfo {author} {\bibfnamefont {S.~X.}\ \bibnamefont
  {Huang}}\ and\ \bibinfo {author} {\bibfnamefont {C.~L.}\ \bibnamefont
  {Chien}},\ }\bibfield  {title} {\enquote {\bibinfo {title} {Extended skyrmion
  phase in epitaxial fege(111) thin films},}\ }\href@noop {} {\bibfield
  {journal} {\bibinfo  {journal} {Phys. Rev. Lett.}\ }\textbf {\bibinfo
  {volume} {108}},\ \bibinfo {pages} {267201} (\bibinfo {year}
  {2012})}\BibitemShut {NoStop}%
\bibitem [{\citenamefont {Woo}\ \emph {et~al.}(2016)\citenamefont {Woo},
  \citenamefont {Litzius}, \citenamefont {Kruger}, \citenamefont {Im},
  \citenamefont {Caretta}, \citenamefont {Richter}, \citenamefont {Mann},
  \citenamefont {Krone}, \citenamefont {Reeve}, \citenamefont {Weigand},
  \citenamefont {Agrawal}, \citenamefont {Lemesh}, \citenamefont {Mawass},
  \citenamefont {Fischer}, \citenamefont {Klaui},\ and\ \citenamefont
  {Beach}}]{Woo2016}%
  \BibitemOpen
  \bibfield  {author} {\bibinfo {author} {\bibfnamefont {S.}~\bibnamefont
  {Woo}}, \bibinfo {author} {\bibfnamefont {K.}~\bibnamefont {Litzius}},
  \bibinfo {author} {\bibfnamefont {B.}~\bibnamefont {Kruger}}, \bibinfo
  {author} {\bibfnamefont {M.-Y.}\ \bibnamefont {Im}}, \bibinfo {author}
  {\bibfnamefont {L.}~\bibnamefont {Caretta}}, \bibinfo {author} {\bibfnamefont
  {K.}~\bibnamefont {Richter}}, \bibinfo {author} {\bibfnamefont
  {M.}~\bibnamefont {Mann}}, \bibinfo {author} {\bibfnamefont {A.}~\bibnamefont
  {Krone}}, \bibinfo {author} {\bibfnamefont {R.~M.}\ \bibnamefont {Reeve}},
  \bibinfo {author} {\bibfnamefont {M.}~\bibnamefont {Weigand}}, \bibinfo
  {author} {\bibfnamefont {P.}~\bibnamefont {Agrawal}}, \bibinfo {author}
  {\bibfnamefont {I.}~\bibnamefont {Lemesh}}, \bibinfo {author} {\bibfnamefont
  {M.-A.}\ \bibnamefont {Mawass}}, \bibinfo {author} {\bibfnamefont
  {P.}~\bibnamefont {Fischer}}, \bibinfo {author} {\bibfnamefont
  {M.}~\bibnamefont {Klaui}}, \ and\ \bibinfo {author} {\bibfnamefont
  {G.~S.~D.}\ \bibnamefont {Beach}},\ }\bibfield  {title} {\enquote {\bibinfo
  {title} {Observation of room-temperature magnetic skyrmions and their
  current-driven dynamics in ultrathin metallic ferromagnets},}\ }\href@noop {}
  {\bibfield  {journal} {\bibinfo  {journal} {Nat Mater}\ }\textbf {\bibinfo
  {volume} {15}},\ \bibinfo {pages} {501--506} (\bibinfo {year}
  {2016})}\BibitemShut {NoStop}%
\bibitem [{\citenamefont {Li}\ \emph {et~al.}(2019)\citenamefont {Li},
  \citenamefont {Lau}, \citenamefont {De~Graef},\ and\ \citenamefont
  {Sokalski}}]{Li2019}%
  \BibitemOpen
  \bibfield  {author} {\bibinfo {author} {\bibfnamefont {M.}~\bibnamefont
  {Li}}, \bibinfo {author} {\bibfnamefont {D.}~\bibnamefont {Lau}}, \bibinfo
  {author} {\bibfnamefont {M.}~\bibnamefont {De~Graef}}, \ and\ \bibinfo
  {author} {\bibfnamefont {V.}~\bibnamefont {Sokalski}},\ }\bibfield  {title}
  {\enquote {\bibinfo {title} {Lorentz {TEM} investigation of chiral spin
  textures and {N}\'eel {S}kyrmions in asymmetric
  [{P}t/({C}o/{N}i)$_{M}$/{I}r]$_{N}$ multi-layer thin films},}\ }\href
  {\doibase https://doi.org/10.1103/PhysRevMaterials.3.064409} {\bibfield
  {journal} {\bibinfo  {journal} {Phys. Rev. Mater.}\ }\textbf {\bibinfo
  {volume} {3}},\ \bibinfo {pages} {064409} (\bibinfo {year}
  {2019})}\BibitemShut {NoStop}%
\bibitem [{\citenamefont {Kiselev}\ \emph {et~al.}(2011)\citenamefont
  {Kiselev}, \citenamefont {Bogdanov}, \citenamefont {Schäfer},\ and\
  \citenamefont {Röler}}]{Kiselev2011}%
  \BibitemOpen
  \bibfield  {author} {\bibinfo {author} {\bibfnamefont {N.~S.}\ \bibnamefont
  {Kiselev}}, \bibinfo {author} {\bibfnamefont {A.~N.}\ \bibnamefont
  {Bogdanov}}, \bibinfo {author} {\bibfnamefont {R.}~\bibnamefont {Schäfer}},
  \ and\ \bibinfo {author} {\bibfnamefont {U.~K.}\ \bibnamefont {Röler}},\
  }\bibfield  {title} {\enquote {\bibinfo {title} {Chiral skyrmions in thin
  magnetic films: New objects for magnetic storage technologies?}}\ }\href
  {\doibase 10.1088/0022-3727/44/39/392001} {\bibfield  {journal} {\bibinfo
  {journal} {J. Phys. D Appl. Phys.}\ }\textbf {\bibinfo {volume} {44}}
  (\bibinfo {year} {2011}),\ 10.1088/0022-3727/44/39/392001}\BibitemShut
  {NoStop}%
\bibitem [{\citenamefont {Tomasello}\ \emph {et~al.}(2014)\citenamefont
  {Tomasello}, \citenamefont {Martinez}, \citenamefont {Zivieri}, \citenamefont
  {Torres}, \citenamefont {Carpentieri},\ and\ \citenamefont
  {Finocchio}}]{Tomasello2014}%
  \BibitemOpen
  \bibfield  {author} {\bibinfo {author} {\bibfnamefont {R.}~\bibnamefont
  {Tomasello}}, \bibinfo {author} {\bibfnamefont {E.}~\bibnamefont {Martinez}},
  \bibinfo {author} {\bibfnamefont {R.}~\bibnamefont {Zivieri}}, \bibinfo
  {author} {\bibfnamefont {L.}~\bibnamefont {Torres}}, \bibinfo {author}
  {\bibfnamefont {M.}~\bibnamefont {Carpentieri}}, \ and\ \bibinfo {author}
  {\bibfnamefont {G.}~\bibnamefont {Finocchio}},\ }\bibfield  {title} {\enquote
  {\bibinfo {title} {A strategy for the design of skyrmion racetrack
  memories},}\ }\href {\doibase 10.1038/srep06784} {\bibfield  {journal}
  {\bibinfo  {journal} {Sci. Rep.}\ }\textbf {\bibinfo {volume} {4}} (\bibinfo
  {year} {2014}),\ 10.1038/srep06784}\BibitemShut {NoStop}%
\bibitem [{\citenamefont {Huang}\ \emph {et~al.}(2017)\citenamefont {Huang},
  \citenamefont {Kang}, \citenamefont {Zhang}, \citenamefont {Zhou},\ and\
  \citenamefont {Zhao}}]{Huang2017}%
  \BibitemOpen
  \bibfield  {author} {\bibinfo {author} {\bibfnamefont {Y.}~\bibnamefont
  {Huang}}, \bibinfo {author} {\bibfnamefont {W.}~\bibnamefont {Kang}},
  \bibinfo {author} {\bibfnamefont {X.}~\bibnamefont {Zhang}}, \bibinfo
  {author} {\bibfnamefont {Y.}~\bibnamefont {Zhou}}, \ and\ \bibinfo {author}
  {\bibfnamefont {W.}~\bibnamefont {Zhao}},\ }\bibfield  {title} {\enquote
  {\bibinfo {title} {Magnetic skyrmion-based synaptic devices},}\ }\href
  {\doibase 10.1088/1361-6528/aa5838} {\bibfield  {journal} {\bibinfo
  {journal} {Nanotechnology}\ }\textbf {\bibinfo {volume} {28}} (\bibinfo
  {year} {2017}),\ 10.1088/1361-6528/aa5838}\BibitemShut {NoStop}%
\bibitem [{\citenamefont {Song}\ \emph {et~al.}(2020)\citenamefont {Song},
  \citenamefont {Jeong}, \citenamefont {Pan}, \citenamefont {Zhang},
  \citenamefont {Zia}, \citenamefont {Cha}, \citenamefont {Park}, \citenamefont
  {Kim}, \citenamefont {Finizio}, \citenamefont {Chang}, \citenamefont {Zhou},
  \citenamefont {Zhao}, \citenamefont {Kang}, \citenamefont {Ju},\ and\
  \citenamefont {Woo}}]{Song2020}%
  \BibitemOpen
  \bibfield  {author} {\bibinfo {author} {\bibfnamefont {K.~M.}\ \bibnamefont
  {Song}}, \bibinfo {author} {\bibfnamefont {J.-S.}\ \bibnamefont {Jeong}},
  \bibinfo {author} {\bibfnamefont {B.}~\bibnamefont {Pan}}, \bibinfo {author}
  {\bibfnamefont {X.}~\bibnamefont {Zhang}}, \bibinfo {author} {\bibfnamefont
  {J.}~\bibnamefont {Zia}}, \bibinfo {author} {\bibfnamefont {S.}~\bibnamefont
  {Cha}}, \bibinfo {author} {\bibfnamefont {T.-E.}\ \bibnamefont {Park}},
  \bibinfo {author} {\bibfnamefont {K.}~\bibnamefont {Kim}}, \bibinfo {author}
  {\bibfnamefont {J.~R.}\ \bibnamefont {Finizio}}, \bibinfo {author}
  {\bibfnamefont {J.}~\bibnamefont {Chang}}, \bibinfo {author} {\bibfnamefont
  {Y.}~\bibnamefont {Zhou}}, \bibinfo {author} {\bibfnamefont {W.}~\bibnamefont
  {Zhao}}, \bibinfo {author} {\bibfnamefont {W.}~\bibnamefont {Kang}}, \bibinfo
  {author} {\bibfnamefont {H.}~\bibnamefont {Ju}}, \ and\ \bibinfo {author}
  {\bibfnamefont {S.}~\bibnamefont {Woo}},\ }\bibfield  {title} {\enquote
  {\bibinfo {title} {{Skyrmion-based artificial synapses for neuromorphic
  computing}},}\ }\href@noop {} {\bibfield  {journal} {\bibinfo  {journal}
  {Nat. Electron.}\ }\textbf {\bibinfo {volume} {3}},\ \bibinfo {pages}
  {148--155} (\bibinfo {year} {2020})}\BibitemShut {NoStop}%
\bibitem [{\citenamefont {Fert}, \citenamefont {Cros},\ and\ \citenamefont
  {Sampaio}(2013)}]{Fert2013}%
  \BibitemOpen
  \bibfield  {author} {\bibinfo {author} {\bibfnamefont {A.}~\bibnamefont
  {Fert}}, \bibinfo {author} {\bibfnamefont {V.}~\bibnamefont {Cros}}, \ and\
  \bibinfo {author} {\bibfnamefont {J.}~\bibnamefont {Sampaio}},\ }\bibfield
  {title} {\enquote {\bibinfo {title} {Skyrmions on the track},}\ }\href@noop
  {} {\bibfield  {journal} {\bibinfo  {journal} {Nat. Nanotechnol.}\ }\textbf
  {\bibinfo {volume} {8}},\ \bibinfo {pages} {152–156} (\bibinfo {year}
  {2013})}\BibitemShut {NoStop}%
\bibitem [{\citenamefont {Juge}\ \emph {et~al.}(2019)\citenamefont {Juge},
  \citenamefont {Je}, \citenamefont {Chaves}, \citenamefont {Buda-Prejbeanu},
  \citenamefont {Peña-Garcia}, \citenamefont {Nath}, \citenamefont {Miron},
  \citenamefont {Rana}, \citenamefont {Aballe}, \citenamefont {Foerster},
  \citenamefont {Genuzio}, \citenamefont {Menteş}, \citenamefont {Locatelli},
  \citenamefont {Maccherozzi}, \citenamefont {Dhesi}, \citenamefont
  {Belmeguenai}, \citenamefont {Roussigné}, \citenamefont {Auffret},
  \citenamefont {Pizzini}, \citenamefont {Gaudin}, \citenamefont {Vogel},\ and\
  \citenamefont {Boulle}}]{Juge2019}%
  \BibitemOpen
  \bibfield  {author} {\bibinfo {author} {\bibfnamefont {R.}~\bibnamefont
  {Juge}}, \bibinfo {author} {\bibfnamefont {S.~G.}\ \bibnamefont {Je}},
  \bibinfo {author} {\bibfnamefont {D.~D.~S.}\ \bibnamefont {Chaves}}, \bibinfo
  {author} {\bibfnamefont {L.~D.}\ \bibnamefont {Buda-Prejbeanu}}, \bibinfo
  {author} {\bibfnamefont {J.}~\bibnamefont {Peña-Garcia}}, \bibinfo {author}
  {\bibfnamefont {J.}~\bibnamefont {Nath}}, \bibinfo {author} {\bibfnamefont
  {I.~M.}\ \bibnamefont {Miron}}, \bibinfo {author} {\bibfnamefont {K.~G.}\
  \bibnamefont {Rana}}, \bibinfo {author} {\bibfnamefont {L.}~\bibnamefont
  {Aballe}}, \bibinfo {author} {\bibfnamefont {M.}~\bibnamefont {Foerster}},
  \bibinfo {author} {\bibfnamefont {F.}~\bibnamefont {Genuzio}}, \bibinfo
  {author} {\bibfnamefont {T.~O.}\ \bibnamefont {Menteş}}, \bibinfo {author}
  {\bibfnamefont {A.}~\bibnamefont {Locatelli}}, \bibinfo {author}
  {\bibfnamefont {F.}~\bibnamefont {Maccherozzi}}, \bibinfo {author}
  {\bibfnamefont {S.~S.}\ \bibnamefont {Dhesi}}, \bibinfo {author}
  {\bibfnamefont {M.}~\bibnamefont {Belmeguenai}}, \bibinfo {author}
  {\bibfnamefont {Y.}~\bibnamefont {Roussigné}}, \bibinfo {author}
  {\bibfnamefont {S.}~\bibnamefont {Auffret}}, \bibinfo {author} {\bibfnamefont
  {S.}~\bibnamefont {Pizzini}}, \bibinfo {author} {\bibfnamefont
  {G.}~\bibnamefont {Gaudin}}, \bibinfo {author} {\bibfnamefont
  {J.}~\bibnamefont {Vogel}}, \ and\ \bibinfo {author} {\bibfnamefont
  {O.}~\bibnamefont {Boulle}},\ }\bibfield  {title} {\enquote {\bibinfo {title}
  {Current-driven skyrmion dynamics and drive-dependent skyrmion hall effect in
  an ultrathin film},}\ }\href {\doibase 10.1103/PhysRevApplied.12.044007}
  {\bibfield  {journal} {\bibinfo  {journal} {Phys. Rev. Appl.}\ }\textbf
  {\bibinfo {volume} {12}} (\bibinfo {year} {2019}),\
  10.1103/PhysRevApplied.12.044007}\BibitemShut {NoStop}%
\bibitem [{\citenamefont {Nagaosa}\ and\ \citenamefont
  {Tokura}(2013)}]{Nagaosa2013}%
  \BibitemOpen
  \bibfield  {author} {\bibinfo {author} {\bibfnamefont {N.}~\bibnamefont
  {Nagaosa}}\ and\ \bibinfo {author} {\bibfnamefont {Y.}~\bibnamefont
  {Tokura}},\ }\bibfield  {title} {\enquote {\bibinfo {title} {Topological
  properties and dynamics of magnetic skyrmions},}\ }\href {\doibase
  10.1038/nnano.2013.243} {\bibfield  {journal} {\bibinfo  {journal} {Nat.
  Nanotechnol.}\ }\textbf {\bibinfo {volume} {8}},\ \bibinfo {pages} {899--911}
  (\bibinfo {year} {2013})}\BibitemShut {NoStop}%
\bibitem [{\citenamefont {You}\ \emph {et~al.}(2012)\citenamefont {You},
  \citenamefont {Sousa}, \citenamefont {Bandiera}, \citenamefont {Rodmacq},\
  and\ \citenamefont {Dieny}}]{You2012}%
  \BibitemOpen
  \bibfield  {author} {\bibinfo {author} {\bibfnamefont {L.}~\bibnamefont
  {You}}, \bibinfo {author} {\bibfnamefont {R.~C.}\ \bibnamefont {Sousa}},
  \bibinfo {author} {\bibfnamefont {S.}~\bibnamefont {Bandiera}}, \bibinfo
  {author} {\bibfnamefont {B.}~\bibnamefont {Rodmacq}}, \ and\ \bibinfo
  {author} {\bibfnamefont {B.}~\bibnamefont {Dieny}},\ }\bibfield  {title}
  {\enquote {\bibinfo {title} {Co/ni multilayers with perpendicular anisotropy
  for spintronic device applications},}\ }\href {\doibase 10.1063/1.4704184}
  {\bibfield  {journal} {\bibinfo  {journal} {Appl. Phys. Lett.}\ }\textbf
  {\bibinfo {volume} {100}} (\bibinfo {year} {2012}),\
  10.1063/1.4704184}\BibitemShut {NoStop}%
\bibitem [{\citenamefont {Rai}\ \emph {et~al.}(2020)\citenamefont {Rai},
  \citenamefont {Sapkota}, \citenamefont {Pokhrel}, \citenamefont {Li},
  \citenamefont {Graef}, \citenamefont {Mewes}, \citenamefont {Sokalski},\ and\
  \citenamefont {Mewes}}]{Rai2020}%
  \BibitemOpen
  \bibfield  {author} {\bibinfo {author} {\bibfnamefont {A.}~\bibnamefont
  {Rai}}, \bibinfo {author} {\bibfnamefont {A.}~\bibnamefont {Sapkota}},
  \bibinfo {author} {\bibfnamefont {A.}~\bibnamefont {Pokhrel}}, \bibinfo
  {author} {\bibfnamefont {M.}~\bibnamefont {Li}}, \bibinfo {author}
  {\bibfnamefont {M.~D.}\ \bibnamefont {Graef}}, \bibinfo {author}
  {\bibfnamefont {C.}~\bibnamefont {Mewes}}, \bibinfo {author} {\bibfnamefont
  {V.}~\bibnamefont {Sokalski}}, \ and\ \bibinfo {author} {\bibfnamefont
  {T.}~\bibnamefont {Mewes}},\ }\bibfield  {title} {\enquote {\bibinfo {title}
  {Higher-order perpendicular magnetic anisotropy and interfacial damping of
  co/ni multilayers},}\ }\href {\doibase 10.1103/PhysRevB.102.174421}
  {\bibfield  {journal} {\bibinfo  {journal} {Phys. Rev. B}\ }\textbf {\bibinfo
  {volume} {102}} (\bibinfo {year} {2020}),\
  10.1103/PhysRevB.102.174421}\BibitemShut {NoStop}%
\bibitem [{\citenamefont {Yang}\ \emph {et~al.}(2015)\citenamefont {Yang},
  \citenamefont {Thiaville}, \citenamefont {Rohart}, \citenamefont {Fert},\
  and\ \citenamefont {Chshiev}}]{Yang2015}%
  \BibitemOpen
  \bibfield  {author} {\bibinfo {author} {\bibfnamefont {H.}~\bibnamefont
  {Yang}}, \bibinfo {author} {\bibfnamefont {A.}~\bibnamefont {Thiaville}},
  \bibinfo {author} {\bibfnamefont {S.}~\bibnamefont {Rohart}}, \bibinfo
  {author} {\bibfnamefont {A.}~\bibnamefont {Fert}}, \ and\ \bibinfo {author}
  {\bibfnamefont {M.}~\bibnamefont {Chshiev}},\ }\bibfield  {title} {\enquote
  {\bibinfo {title} {Anatomy of dzyaloshinskii-moriya interaction at co/pt
  interfaces},}\ }\href {\doibase 10.1103/PhysRevLett.115.267210} {\bibfield
  {journal} {\bibinfo  {journal} {Physical Review Letters}\ }\textbf {\bibinfo
  {volume} {115}} (\bibinfo {year} {2015}),\
  10.1103/PhysRevLett.115.267210}\BibitemShut {NoStop}%
\bibitem [{\citenamefont {Di}\ \emph {et~al.}(2015)\citenamefont {Di},
  \citenamefont {Zhang}, \citenamefont {Lim}, \citenamefont {Ng}, \citenamefont
  {Kuok}, \citenamefont {Yu}, \citenamefont {Yoon}, \citenamefont {Qiu},\ and\
  \citenamefont {Yang}}]{Di2015}%
  \BibitemOpen
  \bibfield  {author} {\bibinfo {author} {\bibfnamefont {K.}~\bibnamefont
  {Di}}, \bibinfo {author} {\bibfnamefont {V.~L.}\ \bibnamefont {Zhang}},
  \bibinfo {author} {\bibfnamefont {H.~S.}\ \bibnamefont {Lim}}, \bibinfo
  {author} {\bibfnamefont {S.~C.}\ \bibnamefont {Ng}}, \bibinfo {author}
  {\bibfnamefont {M.~H.}\ \bibnamefont {Kuok}}, \bibinfo {author}
  {\bibfnamefont {J.}~\bibnamefont {Yu}}, \bibinfo {author} {\bibfnamefont
  {J.}~\bibnamefont {Yoon}}, \bibinfo {author} {\bibfnamefont {X.}~\bibnamefont
  {Qiu}}, \ and\ \bibinfo {author} {\bibfnamefont {H.}~\bibnamefont {Yang}},\
  }\bibfield  {title} {\enquote {\bibinfo {title} {Direct observation of the
  dzyaloshinskii-moriya interaction in a pt/co/ni film},}\ }\href {\doibase
  10.1103/PhysRevLett.114.047201} {\bibfield  {journal} {\bibinfo  {journal}
  {Phys. Rev. Lett.}\ }\textbf {\bibinfo {volume} {114}} (\bibinfo {year}
  {2015}),\ 10.1103/PhysRevLett.114.047201}\BibitemShut {NoStop}%
\bibitem [{\citenamefont {Prudnikov}\ \emph {et~al.}(2018)\citenamefont
  {Prudnikov}, \citenamefont {Li}, \citenamefont {Graef},\ and\ \citenamefont
  {Sokalski}}]{Prudnikov2018}%
  \BibitemOpen
  \bibfield  {author} {\bibinfo {author} {\bibfnamefont {A.}~\bibnamefont
  {Prudnikov}}, \bibinfo {author} {\bibfnamefont {M.}~\bibnamefont {Li}},
  \bibinfo {author} {\bibfnamefont {M.~D.}\ \bibnamefont {Graef}}, \ and\
  \bibinfo {author} {\bibfnamefont {V.}~\bibnamefont {Sokalski}},\ }\bibfield
  {title} {\enquote {\bibinfo {title} {Simultaneous control of interlayer
  exchange coupling and the interfacial {D}zyaloshinskii-{M}oriya interaction
  in {R}u-based synthetic antiferromagnets},}\ }\href@noop {} {\bibfield
  {journal} {\bibinfo  {journal} {IEEE Magn. Lett.}\ }\textbf {\bibinfo
  {volume} {10}},\ \bibinfo {pages} {6100304} (\bibinfo {year}
  {2018})}\BibitemShut {NoStop}%
\bibitem [{\citenamefont {Pandey}\ \emph {et~al.}(2020)\citenamefont {Pandey},
  \citenamefont {Li}, \citenamefont {De~Graef},\ and\ \citenamefont
  {Sokalski}}]{Pandey2020}%
  \BibitemOpen
  \bibfield  {author} {\bibinfo {author} {\bibfnamefont {N.}~\bibnamefont
  {Pandey}}, \bibinfo {author} {\bibfnamefont {M.}~\bibnamefont {Li}}, \bibinfo
  {author} {\bibfnamefont {M.}~\bibnamefont {De~Graef}}, \ and\ \bibinfo
  {author} {\bibfnamefont {V.}~\bibnamefont {Sokalski}},\ }\bibfield  {title}
  {\enquote {\bibinfo {title} {Stabilization of coupled {D}zyaloshinskii domain
  walls in fully compensated synthetic anti-ferromagnets},}\ }\href {\doibase
  https://doi.org/10.1063/1.5130411} {\bibfield  {journal} {\bibinfo  {journal}
  {AIP Adv.}\ }\textbf {\bibinfo {volume} {10}},\ \bibinfo {pages} {015233}
  (\bibinfo {year} {2020})}\BibitemShut {NoStop}%
\bibitem [{\citenamefont {De~Graef}(2000)}]{degraef2000d}%
  \BibitemOpen
  \bibfield  {author} {\bibinfo {author} {\bibfnamefont {M.}~\bibnamefont
  {De~Graef}},\ }\bibfield  {title} {\enquote {\bibinfo {title} {Lorentz
  microscopy: Theoretical basis and image simulations},}\ }in\ \href@noop {}
  {\emph {\bibinfo {booktitle} {Magnetic {M}icroscopy and its {A}pplications to
  {M}agnetic {M}aterials}}},\ \bibinfo {series} {Experimental Methods in the
  Physical Sciences}, Vol.~\bibinfo {volume} {36},\ \bibinfo {editor} {edited
  by\ \bibinfo {editor} {\bibfnamefont {M.}~\bibnamefont {De~Graef}}\ and\
  \bibinfo {editor} {\bibfnamefont {Y.}~\bibnamefont {Zhu}}}\ (\bibinfo
  {publisher} {Academic Press},\ \bibinfo {year} {2000})\ Chap.~\bibinfo
  {chapter} {2}\BibitemShut {NoStop}%
\bibitem [{\citenamefont {Pollard}\ \emph {et~al.}(2017)\citenamefont
  {Pollard}, \citenamefont {Garlow}, \citenamefont {Yu}, \citenamefont {Wang},
  \citenamefont {Zhu},\ and\ \citenamefont {Yang}}]{Pollard2017}%
  \BibitemOpen
  \bibfield  {author} {\bibinfo {author} {\bibfnamefont {S.~D.}\ \bibnamefont
  {Pollard}}, \bibinfo {author} {\bibfnamefont {J.~A.}\ \bibnamefont {Garlow}},
  \bibinfo {author} {\bibfnamefont {J.}~\bibnamefont {Yu}}, \bibinfo {author}
  {\bibfnamefont {Z.}~\bibnamefont {Wang}}, \bibinfo {author} {\bibfnamefont
  {Y.}~\bibnamefont {Zhu}}, \ and\ \bibinfo {author} {\bibfnamefont
  {H.}~\bibnamefont {Yang}},\ }\bibfield  {title} {\enquote {\bibinfo {title}
  {Observation of stable {N}\'eel skyrmions in cobalt/palladium multilayers
  with {L}orentz transmission electron microscopy},}\ }\href@noop {} {\bibfield
   {journal} {\bibinfo  {journal} {Sci. Rep.}\ }\textbf {\bibinfo {volume}
  {8}},\ \bibinfo {pages} {14761} (\bibinfo {year} {2017})}\BibitemShut
  {NoStop}%
\bibitem [{\citenamefont {Lavrijsen}\ \emph {et~al.}(2015)\citenamefont
  {Lavrijsen}, \citenamefont {Hartmann}, \citenamefont {Brink}, \citenamefont
  {Yin}, \citenamefont {Barcones}, \citenamefont {Duine}, \citenamefont
  {Verheijen}, \citenamefont {Swagten},\ and\ \citenamefont
  {Koopmans}}]{Lavrijsen2015}%
  \BibitemOpen
  \bibfield  {author} {\bibinfo {author} {\bibfnamefont {R.}~\bibnamefont
  {Lavrijsen}}, \bibinfo {author} {\bibfnamefont {D.~M.}\ \bibnamefont
  {Hartmann}}, \bibinfo {author} {\bibfnamefont {A.~V.~D.}\ \bibnamefont
  {Brink}}, \bibinfo {author} {\bibfnamefont {Y.}~\bibnamefont {Yin}}, \bibinfo
  {author} {\bibfnamefont {B.}~\bibnamefont {Barcones}}, \bibinfo {author}
  {\bibfnamefont {R.~A.}\ \bibnamefont {Duine}}, \bibinfo {author}
  {\bibfnamefont {M.~A.}\ \bibnamefont {Verheijen}}, \bibinfo {author}
  {\bibfnamefont {H.~J.}\ \bibnamefont {Swagten}}, \ and\ \bibinfo {author}
  {\bibfnamefont {B.}~\bibnamefont {Koopmans}},\ }\bibfield  {title} {\enquote
  {\bibinfo {title} {Asymmetric magnetic bubble expansion under in-plane field
  in pt/co/pt: Effect of interface engineering},}\ }\href {\doibase
  10.1103/PhysRevB.91.104414} {\bibfield  {journal} {\bibinfo  {journal}
  {Physical Review B - Condensed Matter and Materials Physics}\ }\textbf
  {\bibinfo {volume} {91}} (\bibinfo {year} {2015}),\
  10.1103/PhysRevB.91.104414}\BibitemShut {NoStop}%
\bibitem [{\citenamefont {Montoya}\ \emph {et~al.}(2017)\citenamefont
  {Montoya}, \citenamefont {Couture}, \citenamefont {Chess}, \citenamefont
  {Lee}, \citenamefont {Kent}, \citenamefont {Henze}, \citenamefont {Sinha},
  \citenamefont {Im}, \citenamefont {Kevan}, \citenamefont {Fischer},
  \citenamefont {McMorran}, \citenamefont {Lomakin}, \citenamefont {Roy},\ and\
  \citenamefont {Fullerton}}]{Montoya2017}%
  \BibitemOpen
  \bibfield  {author} {\bibinfo {author} {\bibfnamefont {S.~A.}\ \bibnamefont
  {Montoya}}, \bibinfo {author} {\bibfnamefont {S.}~\bibnamefont {Couture}},
  \bibinfo {author} {\bibfnamefont {J.~J.}\ \bibnamefont {Chess}}, \bibinfo
  {author} {\bibfnamefont {J.~C.}\ \bibnamefont {Lee}}, \bibinfo {author}
  {\bibfnamefont {N.}~\bibnamefont {Kent}}, \bibinfo {author} {\bibfnamefont
  {D.}~\bibnamefont {Henze}}, \bibinfo {author} {\bibfnamefont {S.~K.}\
  \bibnamefont {Sinha}}, \bibinfo {author} {\bibfnamefont {M.~Y.}\ \bibnamefont
  {Im}}, \bibinfo {author} {\bibfnamefont {S.~D.}\ \bibnamefont {Kevan}},
  \bibinfo {author} {\bibfnamefont {P.}~\bibnamefont {Fischer}}, \bibinfo
  {author} {\bibfnamefont {B.~J.}\ \bibnamefont {McMorran}}, \bibinfo {author}
  {\bibfnamefont {V.}~\bibnamefont {Lomakin}}, \bibinfo {author} {\bibfnamefont
  {S.}~\bibnamefont {Roy}}, \ and\ \bibinfo {author} {\bibfnamefont {E.~E.}\
  \bibnamefont {Fullerton}},\ }\bibfield  {title} {\enquote {\bibinfo {title}
  {Tailoring magnetic energies to form dipole skyrmions and skyrmion
  lattices},}\ }\href {\doibase 10.1103/PhysRevB.95.024415} {\bibfield
  {journal} {\bibinfo  {journal} {Phys. Rev. B}\ }\textbf {\bibinfo {volume}
  {95}} (\bibinfo {year} {2017}),\ 10.1103/PhysRevB.95.024415}\BibitemShut
  {NoStop}%
\bibitem [{\citenamefont {Je}\ \emph {et~al.}(2020)\citenamefont {Je},
  \citenamefont {Han}, \citenamefont {Kim}, \citenamefont {Montoya},
  \citenamefont {Chao}, \citenamefont {Hong}, \citenamefont {Fullerton},
  \citenamefont {Lee}, \citenamefont {Lee}, \citenamefont {Im},\ and\
  \citenamefont {Hong}}]{Je2020}%
  \BibitemOpen
  \bibfield  {author} {\bibinfo {author} {\bibfnamefont {S.~G.}\ \bibnamefont
  {Je}}, \bibinfo {author} {\bibfnamefont {H.~S.}\ \bibnamefont {Han}},
  \bibinfo {author} {\bibfnamefont {S.~K.}\ \bibnamefont {Kim}}, \bibinfo
  {author} {\bibfnamefont {S.~A.}\ \bibnamefont {Montoya}}, \bibinfo {author}
  {\bibfnamefont {W.}~\bibnamefont {Chao}}, \bibinfo {author} {\bibfnamefont
  {I.~S.}\ \bibnamefont {Hong}}, \bibinfo {author} {\bibfnamefont {E.~E.}\
  \bibnamefont {Fullerton}}, \bibinfo {author} {\bibfnamefont {K.~S.}\
  \bibnamefont {Lee}}, \bibinfo {author} {\bibfnamefont {K.~J.}\ \bibnamefont
  {Lee}}, \bibinfo {author} {\bibfnamefont {M.~Y.}\ \bibnamefont {Im}}, \ and\
  \bibinfo {author} {\bibfnamefont {J.~I.}\ \bibnamefont {Hong}},\ }\bibfield
  {title} {\enquote {\bibinfo {title} {Direct demonstration of topological
  stability of magnetic skyrmions via topology manipulation},}\ }\href
  {\doibase 10.1021/acsnano.9b08699} {\bibfield  {journal} {\bibinfo  {journal}
  {ACS Nano}\ }\textbf {\bibinfo {volume} {14}},\ \bibinfo {pages} {3251--3258}
  (\bibinfo {year} {2020})}\BibitemShut {NoStop}%
\bibitem [{\citenamefont {Zhang}\ \emph {et~al.}(2018)\citenamefont {Zhang},
  \citenamefont {Zhang}, \citenamefont {Wen}, \citenamefont {Chudnovsky},\ and\
  \citenamefont {Zhang}}]{Zhang2018field}%
  \BibitemOpen
  \bibfield  {author} {\bibinfo {author} {\bibfnamefont {S.}~\bibnamefont
  {Zhang}}, \bibinfo {author} {\bibfnamefont {J.}~\bibnamefont {Zhang}},
  \bibinfo {author} {\bibfnamefont {Y.}~\bibnamefont {Wen}}, \bibinfo {author}
  {\bibfnamefont {E.~M.}\ \bibnamefont {Chudnovsky}}, \ and\ \bibinfo {author}
  {\bibfnamefont {X.}~\bibnamefont {Zhang}},\ }\bibfield  {title} {\enquote
  {\bibinfo {title} {Determination of chirality and density control of
  {N}\'eel-type skyrmions with in-plane magnetic field},}\ }\href@noop {}
  {\bibfield  {journal} {\bibinfo  {journal} {Comm. Phys.}\ }\textbf {\bibinfo
  {volume} {1}},\ \bibinfo {pages} {36} (\bibinfo {year} {2018})}\BibitemShut
  {NoStop}%
\bibitem [{\citenamefont {Fallon}\ \emph {et~al.}(2020)\citenamefont {Fallon},
  \citenamefont {McVitie}, \citenamefont {Legrand}, \citenamefont {Ajejas},
  \citenamefont {Maccariello}, \citenamefont {Collin}, \citenamefont {Cros},\
  and\ \citenamefont {Reyren}}]{Fallon2020}%
  \BibitemOpen
  \bibfield  {author} {\bibinfo {author} {\bibfnamefont {K.}~\bibnamefont
  {Fallon}}, \bibinfo {author} {\bibfnamefont {S.}~\bibnamefont {McVitie}},
  \bibinfo {author} {\bibfnamefont {W.}~\bibnamefont {Legrand}}, \bibinfo
  {author} {\bibfnamefont {F.}~\bibnamefont {Ajejas}}, \bibinfo {author}
  {\bibfnamefont {D.}~\bibnamefont {Maccariello}}, \bibinfo {author}
  {\bibfnamefont {S.}~\bibnamefont {Collin}}, \bibinfo {author} {\bibfnamefont
  {V.}~\bibnamefont {Cros}}, \ and\ \bibinfo {author} {\bibfnamefont
  {N.}~\bibnamefont {Reyren}},\ }\bibfield  {title} {\enquote {\bibinfo {title}
  {Quantitative imaging of hybrid chiral spin textures in magnetic multilayer
  systems by {L}orentz microscopy},}\ }\href@noop {} {\bibfield  {journal}
  {\bibinfo  {journal} {Phys. Rev. B}\ }\textbf {\bibinfo {volume} {100}},\
  \bibinfo {pages} {214431} (\bibinfo {year} {2020})}\BibitemShut {NoStop}%
\bibitem [{\citenamefont {Li}\ \emph {et~al.}(2020)\citenamefont {Li},
  \citenamefont {Sapkota}, \citenamefont {Rai}, \citenamefont {Pokhrel},
  \citenamefont {Mewes}, \citenamefont {Mewes}, \citenamefont {Xiao},\ and\
  \citenamefont {Sokalski}}]{Li2020}%
  \BibitemOpen
  \bibfield  {author} {\bibinfo {author} {\bibfnamefont {M.}~\bibnamefont
  {Li}}, \bibinfo {author} {\bibfnamefont {A.}~\bibnamefont {Sapkota}},
  \bibinfo {author} {\bibfnamefont {A.}~\bibnamefont {Rai}}, \bibinfo {author}
  {\bibfnamefont {A.}~\bibnamefont {Pokhrel}}, \bibinfo {author} {\bibfnamefont
  {T.}~\bibnamefont {Mewes}}, \bibinfo {author} {\bibfnamefont
  {C.}~\bibnamefont {Mewes}}, \bibinfo {author} {\bibfnamefont
  {M.}~\bibnamefont {Xiao}, \bibfnamefont {D.~De~Graef}}, \ and\ \bibinfo
  {author} {\bibfnamefont {V.}~\bibnamefont {Sokalski}},\ }\bibfield  {title}
  {\enquote {\bibinfo {title} {Formation of zero-field skyrmion arrays in
  asymmetric superlattices},}\ }\href@noop {} {\bibfield  {journal} {\bibinfo
  {journal} {Appl. Phys. Lett.}\ }\textbf {\bibinfo {volume} {117}},\ \bibinfo
  {pages} {112403} (\bibinfo {year} {2020})}\BibitemShut {NoStop}%
\end{thebibliography}%

\end{document}